\pdfoutput=1

\documentclass[11pt]{article}
\usepackage{amsmath}
\usepackage{amssymb}
\usepackage{mathtools}
\usepackage{amsthm}

\usepackage{booktabs}
\usepackage[final]{acl}

\usepackage{times}
\usepackage{latexsym}

\usepackage[T1]{fontenc}

\usepackage[utf8]{inputenc}

\usepackage{microtype}

\usepackage{inconsolata}

\usepackage{graphicx}

\title{Information Extraction from Visually Rich Documents 
using LLM-based Organization of Documents into Independent Textual Segments}

\author{Aniket Bhattacharyya\textsuperscript{1}, Anurag Tripathi, Ujjal Das, Archan Karmakar, Amit Pathak, Maneesh Gupta \\ 
Amazon\\
{\tt\small \textsuperscript{1}anikettb@amazon.com}
}

\begin{document}
\maketitle
\begin{abstract}
Information extraction (IE) from Visually Rich Documents (VRDs) containing layout features along with text is a critical and well-studied task. Specialized non-LLM NLP-based solutions typically involve training models using both textual and geometric information to label sequences/tokens as named entities or answers to specific questions. However, these approaches lack reasoning, are not able to infer values not explicitly present in documents, and do not generalize well to new formats. Generative LLM-based approaches proposed recently are capable of reasoning, but struggle to comprehend clues from document layout especially in previously unseen document formats, and do not show competitive performance in heterogeneous VRD benchmark datasets. In this paper, we propose BLOCKIE, a novel LLM-based approach that organizes VRDs into localized, reusable semantic textual segments called \textit{semantic blocks}, which are processed independently. Through focused and more generalizable reasoning,our approach outperforms the state-of-the-art on public VRD benchmarks by 1-3\% in F1 scores, is resilient to document formats previously not encountered and shows abilities to correctly extract information not explicitly present in documents.
\end{abstract}

\section{Introduction}
\label{submission}
\begin{figure*}[ht]
\vskip 0.2in
\begin{center}
\centerline{\includegraphics[width=\columnwidth]{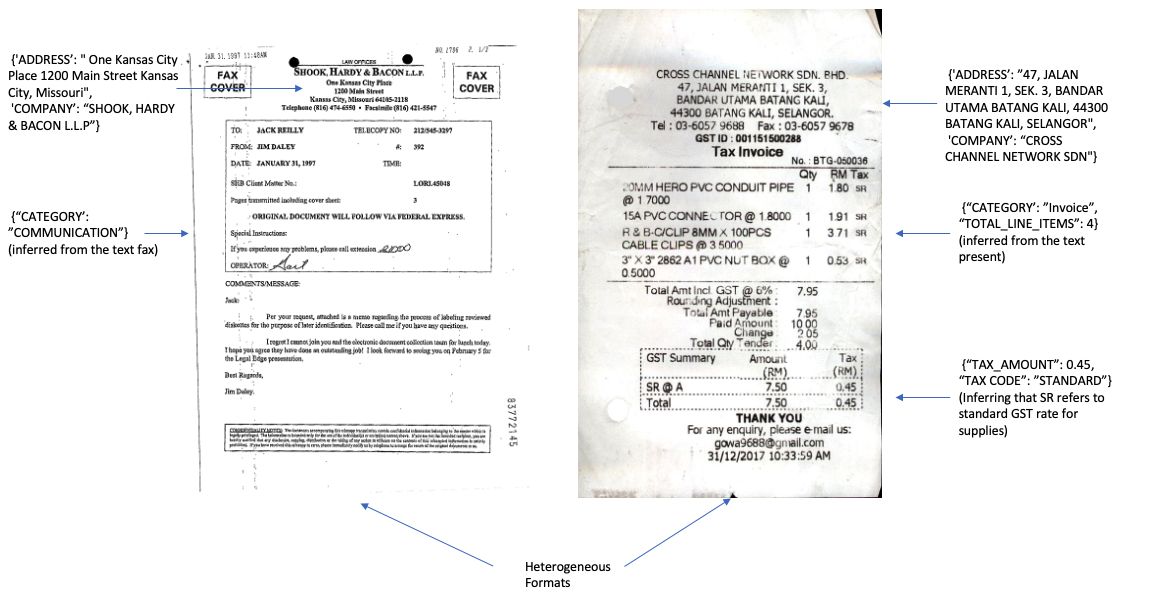}}
\caption{The Information Extraction Task, illustrated using sample images from \cite{funsd} and \cite{Huang_2019}}
\label{task}
\end{center}
\vskip -0.2in
\end{figure*}
Visually Rich Document Understanding (VRDU) is a well researched topic due to its wide industry applicability. Structured or semi-structured documents such as invoices, forms, contracts, receipts etc are handled by most organizations, and for large organizations the volume of such documents can be massive. Processing these documents, especially those of a financial or legal nature, is vital. Figure \ref{task} shows a typical application of VRDU. As can be seen, an ideal information extraction or processing solution, should have the following desiderata - 
\begin{itemize}
    \item High-quality extraction - High precision and recall of desired entities (such as company name or address) to be extracted.
    \item Handling heterogeneity of formats and languages - Handling documents from various sources with different templates (legal fax from US and supplies store invoice from Indonesia in Figure \ref{task}). Public datasets such as \citealp{iitcdip} illustrate the degree of heterogeneity found in real life applications.
    \item Handling new document formats - Solution should be able to handle documents with formats not seen during its training to avoid failure in production environment.
    \item Ability to perform value-absent inference - Entities to be extracted (such as number of line items in Figure \ref{task}) may not always be present explicitly, and may need to be inferred.
\end{itemize} 

A typical approach to document information extraction begins with Optical Character Recognition (OCR) using tools like Amazon Textract or Tesseract \cite{tesstract}. However, OCR alone fails to address several key challenges. Documents exhibit diverse formats and structures, requiring spatial reasoning to correctly associate text with their semantic roles. Systems must understand contextual relationships - for instance, recognizing that 'CGST', 'VAT', and 'SR' all represent tax types, or identifying a vendor name without explicit labels. Additionally, solutions must generalize across heterogeneous document layouts and languages.

Recent approaches have attempted to address these challenges through layout-aware NLP models \citealp{layoutlm,layoutlmv3,peng2022ernielayout, cvpr2023geolayoutlm} enhance text processing with spatial information through mechanisms using cross-attention between text and bounding box embeddings. While effective for template-matching, we show that these models struggle with generalizing to new document formats, making inferences about implicit or absent values, and understanding semantic relationships beyond training examples.

Large Language Models have demonstrated strong reasoning capabilities through chain-of-thought demonstrations \cite{wei2023chainofthought} and few-shot examples attached to the prompt \cite{brown2020languagemodelsfewshotlearners}. However, LLMs face their own limitations: they struggle with processing documents dissimilar to few-shot examples, handling complex layouts efficiently, and scaling prompts for multiple entity extraction. Even approaches using dynamic example selection based on document similarity \cite{perot2024lmdxlanguagemodelbaseddocument} require at least one document with matching format in the labeled sample.

In this work, we propose BLOCKIE, a novel information extraction algorithm that leverages \textit{semantic block}-level parsing. Our approach first identifies self-contained groups of text tokens (\textit{semantic blocks}) and processes them using LLM-driven reasoning informed by similar blocks from labeled samples (see Figure \ref{blocks} for an example on how documents with different templates can have similar blocks). Since semi-structured documents naturally organize information in human-readable blocks (Figure \ref{diff_docs_same_block}), this localized reasoning generalizes well across different document formats. BLOCKIE mimics human document processing by first understanding local regions (\textbf{B}lock \textbf{L}evel \textbf{O}rganization) and then leveraging \textbf{C}ontextual \textbf{K}nowledge from other blocks to stitch information together for IE.

We show that our approach outperforms the state-of-the-art on public benchmark datasets and satisfies all the desiderata for an IE solution.
To summarize, we make the following contributions:
\begin{itemize}
\item We introduce BLOCKIE: Block-Level Organization and Contextual Knowledge-based Information Extraction, a novel algorithm for VRDU that organizes documents into self-contained segments of text tokens called semantic blocks, which are processed using reasoning that generalizes across document formats.

\item We apply BLOCKIE to public benchmark datasets CORD, FUNSD and SROIE, and show that our method concurrently outperforms the current state-of-the-art on all these three datasets by 1-3\% in F1 score.

\item We show that block-level reasoning makes BLOCKIE robust to heterogeneous document databases and new document formats, prevents degradation of performance with smaller LLMs, and allows LLMs to perform value-absent inference.
\end{itemize}

\section{Related Work}
\label{Related Work}
Prior work in VRD understanding can be broadly categorized into three approaches: traditional methods, layout-aware models, and large language models. We discuss each in turn, highlighting their capabilities and limitations.

\textbf{Traditional Methods} initially relied on rule-based systems and handcrafted features \cite{Gorman,jakha,Simon,Marinai,mausam,chiticariu}. While these approaches worked for known templates, they failed to generalize to new document formats. Later deep learning approaches leveraged RNNs \cite{aggarwal-etal-2020-form2seq,palm}, CNNs \cite{HaoCNN,bertgrid,chargrid}, and transformers \cite{wang,majumder} to extract structural information from documents. However, these methods required extensive component-level labeling, limiting their practical applicability.

\textbf{Layout-aware NLP Models} enhanced traditional approaches by incorporating document layout information. Several architectural innovations were proposed: transformers \cite{appalaraju2021docformer,hwang2021spatial,bai2022wukongreader,dhouib2023docparser} for spatial understanding,  layout-aware language models combining BERT-style architectures \cite{devlin2019bert,liu2019roberta,bao2020unilmv2} with spatial information through learnable modules, 2D position embeddings \cite{layoutlm}, and attention mechanisms \cite{layoutlmv2,layoutlmv3,peng2022ernielayout}. Further advances introduced geometric pre-training \cite{cvpr2023geolayoutlm}, graph contrastive learning \cite{formnetv2}, and unified frameworks for simultaneous text detection and classification \cite{yang2023modelingentitiessemanticpoints}. Recent work has improved these models through reading-order prediction \cite{zhang2024modelinglayoutreadingorder}. While these approaches achieve strong performance when fine-tuned on benchmark datasets like DocVQA \cite{docvqa} and FUNSD \cite{funsd} after pre-training on large document corpora like IIT-CDIP \cite{iitcdip}, they remain limited by their token-classification approach, requiring explicit answer presence and struggling with new document formats.

\textbf{Large Language Models} represent the newest approach to VRD understanding. Commercial models like Claude \cite{Anthropic} and ChatGPT \cite{OpenAI} demonstrate zero-shot reasoning capabilities, with Claude 3 achieving state-of-the-art performance on DocVQA \cite{Anthropicres}. Open-source models like LLaVa \cite{llava} and CogVLM \cite{cogvlm} show promise on visual question answering tasks but struggle with zero-shot and multi-entity extraction \cite{bhattacharyya2024informationextractionheterogeneousdocuments}.

Recent work has explored specialized LLM applications for information extraction, particularly in Named Entity Recognition \cite{keraghel2024recentadvancesnamedentity,laskar2023systematicstudycomprehensiveevaluation,ashok2023promptnerpromptingnamedentity,wang2023gptnernamedentityrecognition}. For VRD-specific challenges, researchers have developed layout-aware pre-training \cite{luo2024layoutllmlayoutinstructiontuning}, disentangled spatial attention \cite{wang2023docllmlayoutawaregenerativelanguage}, and normalized line-level bounding box representations \cite{perot2024lmdxlanguagemodelbaseddocument}. However, these approaches have yet to surpass layout-aware NLP methods, and attempts to convert generative models to token-labeling systems often sacrifice their inference capabilities.
\section{Semantic Blocks in VRDs}
\label{background}
\begin{figure}[ht]
\vskip 0.2in
\begin{center}
\centerline{\includegraphics[width=\columnwidth]{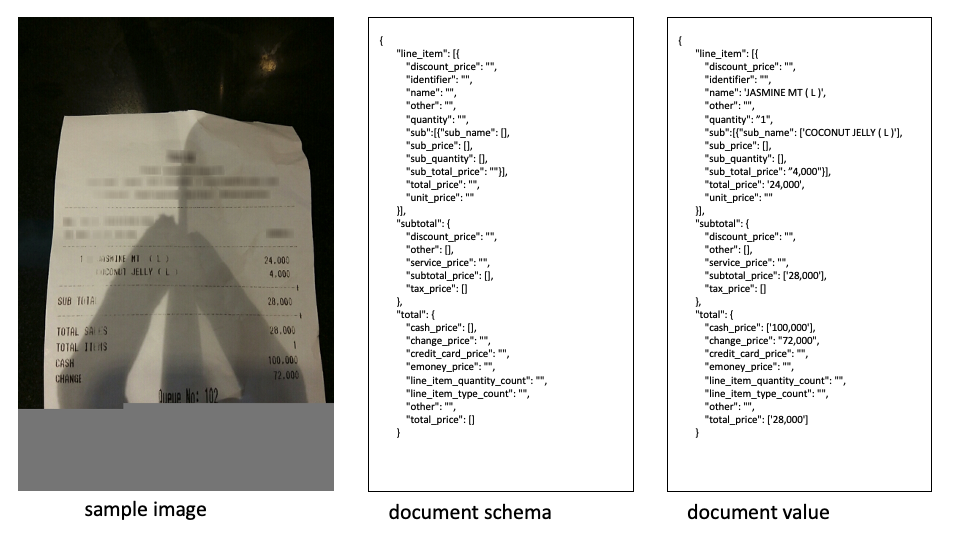}}
\caption{Sample image with document schema and value}
\label{fig:partial}
\end{center}
\vskip -0.2in
\end{figure}

In this section, we define the concept of semantic blocks, and show how these are created practically.

Let us consider a set of documents $\mathcal{D}$ with a common set of hierarchical entities of interest $\mathbb{E}$, which we refer to as the document schema. Let $\mathcal{V}$ denote the set of all possible instantiations of $\mathbb{E}$. Given a document $D \in \mathcal{D}$, let $V_\mathbb{E}(D)\in \mathcal{V}$ denote the actual values of the entities $\mathbb{E}$ for $D$ (for reference, consider sample document, schema and value in Figure \ref{fig:partial}).

For a document $D \in \mathcal{D}$, let $\mathcal{B_D}$ denote the set of all possible segments (i.e. localized visual regions) of D. For any segment $B \in \mathcal{B_D}$, let $V_{\mathbb{E}}(B)$ represent the document values with only entities present in B populated, other entities being blank. Note that $D \in \mathcal{B_D}$ is a special segment comprising of the entire document.

\begin{figure}[ht]
\vskip 0.2in
\begin{center}
\centerline{\includegraphics[width=\columnwidth]{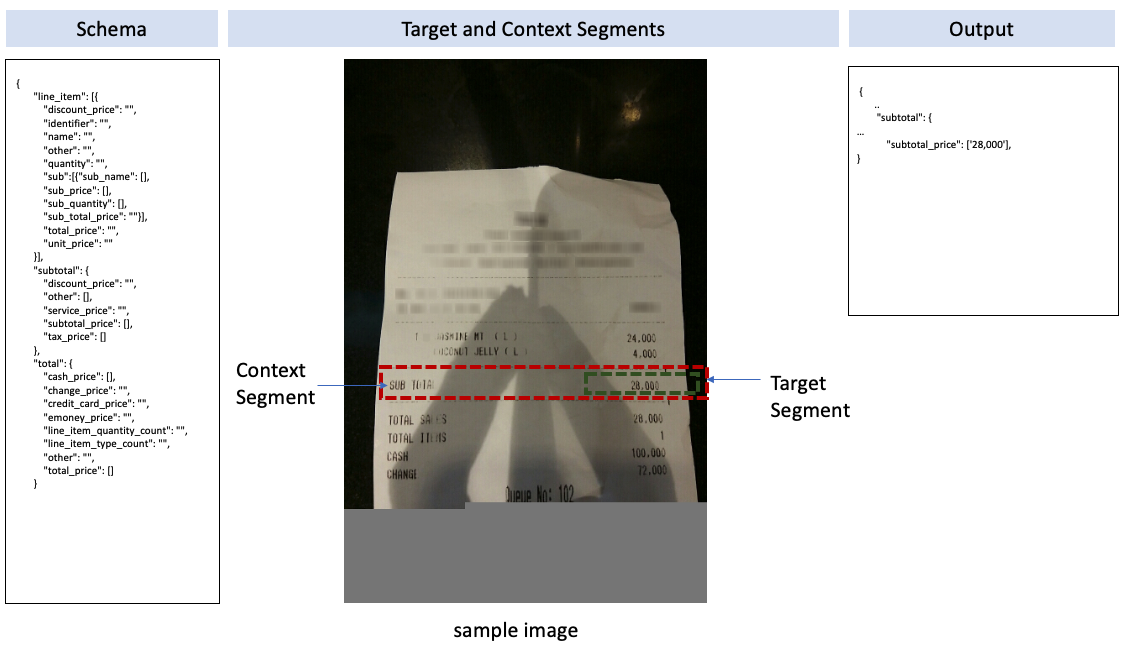}}
\caption{Sample image with document schema and value}
\label{fig:annotation}
\end{center}
\vskip -0.2in
\end{figure}

The annotation operation can be thought of as an attempt to map a segment of a document to the document schema. As input, it takes in the target document segment, and parses it in the context of a larger segment with respect to the schema. The context segment could be any superset of the target, including (typically) the target segment itself or the entire document. Figure \ref{fig:annotation} illustrates the annotation operation with a target and context segment.

Formally, for a given document schema $\mathbb{E}$, the annotation operation can be defined as a mapping $v:\mathcal{B_D} \times \mathcal{B_D} \mapsto \mathcal{V}$. If the annotation is correct, we have,
\begin{equation}
v(B,D) = V_{\mathbb{E}}(B), \forall B \in \mathbb{B_D},  \forall D \in \mathbb{D} 
\label{eq:annotation}
\end{equation}

Now, consider any segment $B \in \mathcal{B_D}$ for a  $D \in \mathcal{D}$. We define B as a \textit{semantic block} if and only if:

\begin{equation}
v(B,B) = v(B,D) = V_\mathbb{E}(B)
\label{eq:semantic_block}
\end{equation}

In other words, a \textit{semantic block} must be interpretable independently without any additional context - the values extracted from B in isolation must match those extracted with full document context.

To illustrate, consider Figure \ref{fig:partial}. In this example, $B_1$: (SUB TOTAL 28.000) is a semantic block with:
\begin{equation*}
v(B_1,D)=subtotal:\{subtotal\_price:[28.000]\}
\end{equation*}
and $B_2$: (TOTAL SALE 28.0000) is a semantic block with:
\begin{equation*}
v(B_2,D) = total:\{total\_price:[28.000]\}
\end{equation*}

On the other hand, (COCONUT JELLY ( L ), 4.000) cannot be a semantic block, as without the context of (1 JASMINE MT (L) 24.000), it is not possible to determine whether it is a sub-item and, if so, which line item it is a sub-item of.

Now, to create semantic blocks in practice, we introduce the concept of semantic \textit{atoms} - the fundamental units for information extraction from VRDs. A semantic atom is an indivisible visual region containing text that forms a complete semantic unit while maintaining spatial coherence through proximity as well as horizontal or vertical alignment. The key characteristic of a semantic atom is that it cannot be decomposed further without losing its intended meaning. For example, in Figure \ref{fig:partial}, ``TOTAL ITEMS'' forms a semantic atom because splitting it into ``TOTAL'' and ``ITEMS'' individually would lose the specific meaning of `number of items' - ``TOTAL'' alone could refer to price or quantity, while ``ITEMS'' alone loses specificity. Moreover, these words maintain spatial coherence through horizontal proximity in the document. Conversely, ``TOTAL ITEMS 1'', although coherent semantically and linked as an attribute value pair, is not spatially proximate, and hence is not an atom, but makes up two linked semantic atoms.

Note that there could be two different types of linkages between semantic atoms in a VRD - linkages of the form attribute:value, or linkages of hierarchy. By hierarchically linked semantic atoms we refer to semantic atoms that belong to hierarchical entities in the document schema. In practice, semantic blocks are \textit{collections of semantic atoms}, such that \textit{all linkages for each atom in the collection is present inside the collection itself}.  This is a sufficient condition for equation \ref{eq:semantic_block}, as given a schema, all context needed to parse any group of atoms is present in a collection of atoms linked to it as hierarchically or as attribute-value. To continue the example, (TOTAL SALE 28.0000) and (SUB TOTAL 28.000) are linked semantic atoms, and (1 JASMINE MT (L) 24.000 COCONUT JELLY ( L ), 4.000) are linked semantic atoms. 

This theoretical foundation guides our development of practical algorithms for document processing, as we will demonstrate in subsequent sections. By decomposing documents into smaller, more generalizable semantic blocks, we can better handle the complexities of varying layouts while maintaining the semantic relationships crucial for accurate information extraction.
\section{Proposed Methodology}
\label{proposed}
\begin{figure}[ht]
\vskip 0.2in
\begin{center}
\centerline{\includegraphics[width=\columnwidth]{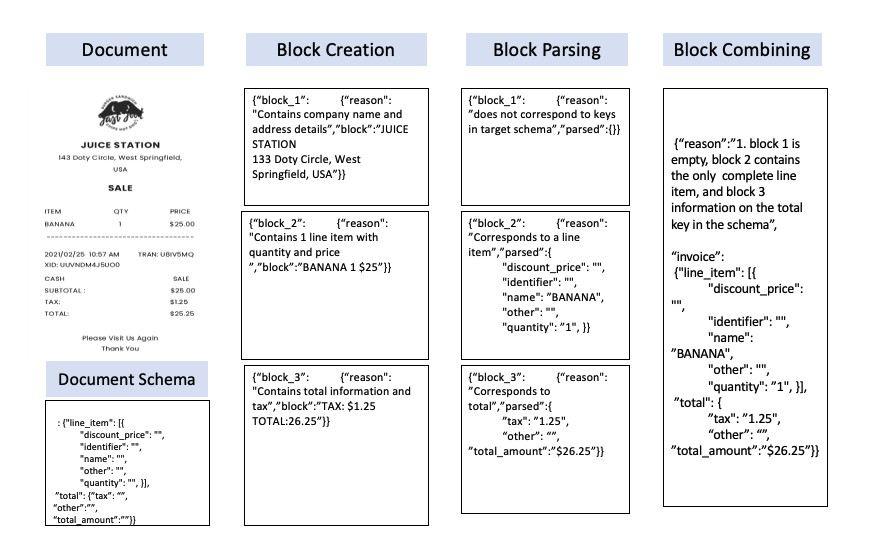}}
\caption{Illustrative flow with a simulated receipt and schema resembling CORD output requirement. The schema is passed along with the output of the block creator along with parses of similar blocks to block parser. Parsed blocks with target schema are then passed to get final output. Reasons are output at each stage.}
\label{flow}
\end{center}
\vskip -0.2in
\end{figure}

Given a group of documents and a required set of entities that need to be extracted in the form of document schema,  we first divide the document into a collection of semantic blocks of related text. These blocks are then processed, which allows LLMs to develop generalizable abstract rules for IE. These partial block parses are then combined to return the set of entities required. However, prior to these steps, it is necessary to convert the train dataset labels to appropriate format, i.e. to independent blocks and their annotations. Further details on each of these steps are provided below.

\textbf{Train Dataset Labelling}
The train dataset is used as a labelled sample. VRD benchmarks such as \citealp{park2019cord} 
generally contain ground truth labels in a key-value format, with appropriate hierarchy and linkages. These are passed to an LLM along with document schema to return three things - (1) step-by-step reasoning for choosing a segment as a block (i.e. self-contained segments of linked atoms, as defined in section \ref{background}), (2) the words in the block, and (3) the partial annotation of the block, using the ground truth labels. All of these three outputs are used downstream.

\subsection{Block Creation}
\label{submission}
\begin{figure}[ht]
\vskip 0.2in
\begin{center}
\centerline{\includegraphics[width=\columnwidth]{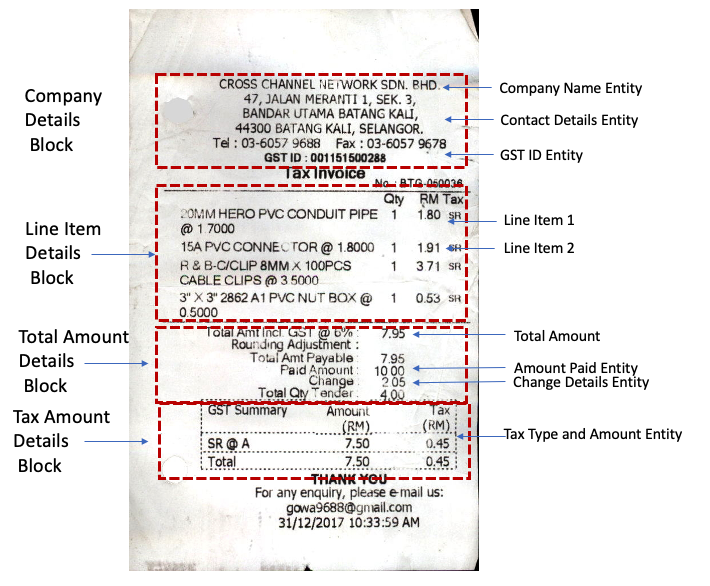}}
\caption{Motivating example for the conceptualization of VRD IE as the parsing of related semantic entities organized in blocks. The entities within a block are related which allows a human to understand that the address in the company details block belongs to the invoicing company instead of say the customer.}
\label{blocks}
\end{center}
\vskip -0.2in
\end{figure}

Given a document from the test dataset, we prompt the LLM to create blocks using the document schema, OCR text and bounding boxes, and dynamic few-shot examples from the labelled train dataset using cosine similarity of OCR text\footnote{\citealp{perot2024lmdxlanguagemodelbaseddocument} show that using similar documents in in-context learning examples improves performance in VRDs.}.The LLM leverages the step by step reasoning from the train dataset blocks on the few-shot samples to understand when a text segment can be considered a block. Note that while we used OCR text and bounding boxes, for multimodal LLMs one can pass the image directly. The creation of self-contained blocks is crucial; in section \ref{Experimental Setup and Results}, we evaluate the impact of block creation on overall accuracy.

\subsection{Block Parsing}
\label{parsing}
\begin{figure}[ht]
\vskip 0.2in
\begin{center}
\centerline{\includegraphics[width=\columnwidth]{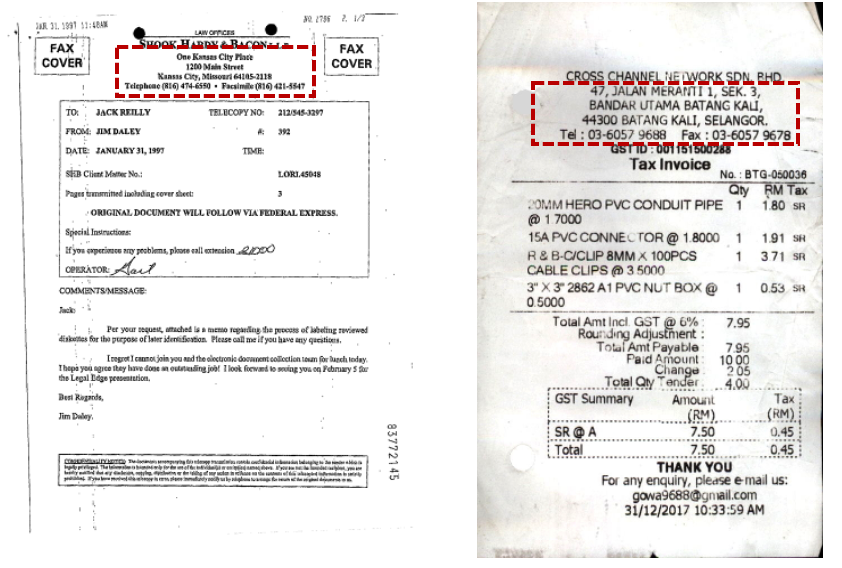}}
\caption{Two documents with different formats (a fax from a legal firm and a supplies store invoice) sharing a similar semantic block corresponding to contact information}
\label{diff_docs_same_block}
\end{center}
\vskip -0.2in
\end{figure}

Once blocks have been created, these are annotated by block parsers. As shown in figure  \ref{diff_docs_same_block}, similar semantically meaningful blocks are found even in documents with different formats. Since these blocks are self-contained, they can be parsed independently.

The document schema is passed to the LLM with few-shot examples of the most similar blocks. The step-by-step reasoning of train dataset block parser triggers similar reasoning in the block parser, and the document schema guides it to return structured output in required format.

\begin{figure}[ht]
\vskip 0.2in
\begin{center}
\centerline{\includegraphics[width=\columnwidth]{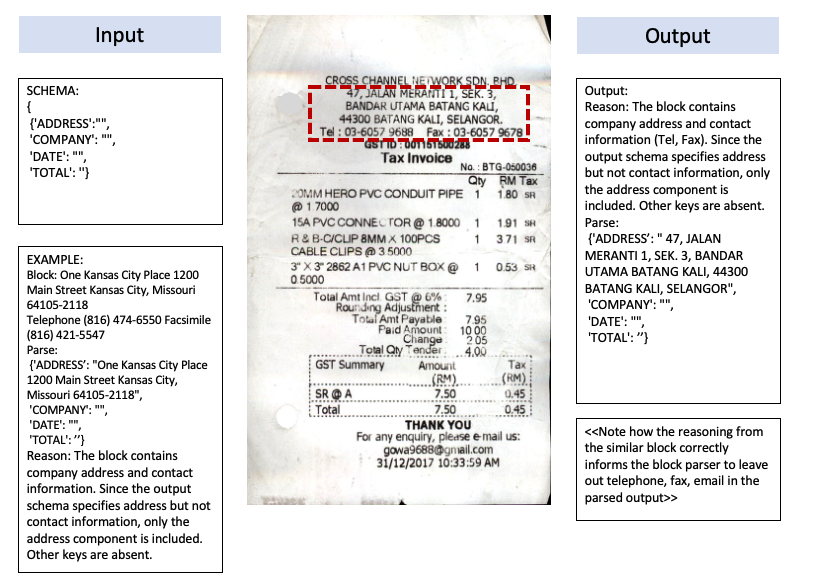}}
\caption{Block Parser on Figure \ref{diff_docs_same_block}, where the legal firm fax is used as a labelled train dataset example, and the supplies store invoice is treated as a test sample}
\label{fig:block_parser}
\end{center}
\vskip -0.2in
\end{figure}

Figure \ref{fig:block_parser} shows how the same example with similar blocks would be annotated by the block parser.

\subsection{Combining Blocks}
Finally, the document schema, blocks and their parses are provided to LLMs to return the entire filled out schema. The LLM acts as a judge assessing the block-parsing reason from the previous steps to stitch together the filled out document schema. Each semantic block benefits from being compared with similar blocks in other documents (which may be heterogenous), and the document schema guides the llm to return structured output.

Figure \ref{flow} illustrates these three steps using a sample document and schema.
Detailed prompts are provided in the appendix.

\section{Experimental Setup and Results} 
\label{Experimental Setup and Results}
We designed our experimental evaluation to rigorously assess BLOCKIE's effectiveness in addressing these challenges. Our analysis examines the method stands up against the desiderata for an ideal information extraction solution for a large heterogenous document database.
\subsection{Experimental Setup}

We evaluate BLOCKIE on three established information extraction benchmarks: CORD \citep{park2019cord}, which focuses on restaurant receipts with hierarchical field structures; FUNSD \citep{funsd}, a subset of \citealp{harley2015icdar}; and SROIE \citep{Huang_2019}, a receipt information extraction dataset. For FUNSD, we focus on entity linking as the original semantic entity classifications (question, answer, header, others) are not meaningful and do not align with real-world information extraction requirements.

To assess the generality of our approach, we conduct experiments for BLOCKIE with multiple language models of varying parameter counts: Claude 3.5 Sonnet \citep{Anthropic35} and four variants of Qwen 2.5 \citep{qwen2025qwen25technicalreport} with 7B, 14B, 32B, and 72B parameters respectively. We used 5 few shot-examples in the prompts for both block creator and parser. Following standard practice in document information extraction, we use the F1 score as our primary evaluation metric. For performance comparison, we consider state-of-the-art methods discussed in section \ref{Related Work}, and we also conduct additional experiments with LayoutLMV3 \citep{layoutlmv3} to show the limitations of layout-aware NLP methods. Additional details about the datasets and implementations are present in Appendix \ref{sec:appendix2}.

\subsection{Results}
\subsubsection{Performance Analysis}

Table \ref{tab:performance} presents BLOCKIE's performance compared to existing approaches across all three datasets. Using Sonnet as the base LLM, BLOCKIE achieves state-of-the-art performance, surpassing both traditional layout-aware approaches and recent LLM-based methods. Notably, BLOCKIE achieves 98.83\% F1-score on CORD, 92.15\% on FUNSD, and 98.52\% on SROIE, establishing new benchmarks across all datasets.
To verify that these improvements stem from our block-based methodology rather than just LLM capabilities, we compare against zero-shot and few-shot variants of Sonnet. The performance gap between BLOCKIE and these baseline approaches (shown in Table \ref{tab:performance}) demonstrates that the improvements arise from our semantic block methodology rather than raw LLM capabilities.
\begin{table*}[t]
\vskip 0.15in
\begin{small}
\begin{tabular}{lllll}
\toprule
Approach      & Method                              & FUNSD  & CORD   & SROIE  \\
\midrule
      &                                     & EL     & SER    & SER    \\
\midrule
      & DocTr\citep{feng2022doctrdocumentimagetransformer}153M           & 73.9   & 98.2   & -      \\
      & LayoutLMv3\citep{layoutlmv3}368M     & 79.37 & 96.98 & 96.12 \\

      & DocFormer\citep{appalaraju2021docformer} 502M & -      & 96.99  & -      \\
Layout-Aware NLP      & FormNet\citet{formnetv2} large         & -      & 97.28  & -      \\

& ERNIE-Layout\citep{peng2022ernielayout}large   & -      & 97.21  & 97.55  \\

 & GeoLayoutLM\citep{cvpr2023geolayoutlm}399M      & 88.06  & 98.11 & 96.62 \\
  & ESP\citep{yang2023modelingentitiessemanticpoints}50M              & 88.88 & 95.65 & -      \\
      & RORE-GeoLayoutLM \citep{zhang2024modelinglayoutreadingorder} 399M+24             & 88.46  & 98.52  & 96.97  \\
      \midrule
  & DocLLM\citep{wang2023docllmlayoutawaregenerativelanguage}                           & -      & 67.4   & 91.9   \\
      & LMDX-Gemini Pro\citep{perot2024lmdxlanguagemodelbaseddocument}                     & -      & 95.57  &        \\
LLM             & LayoutLLM\citep{luo2024layoutllmlayoutinstructiontuning}                         & -      & 63.1   & 72.72  \\ & Sonnet - Zero shot                           & -     & 88.92   & 91.37  \\ & Sonnet - Few shot                           & -      & 95.72   & 96.72  \\
      \midrule
Ours          & BLOCKIE - Sonnet                    & \textbf{92.15}  & \textbf{98.83}  & \textbf{98.52 }
\end{tabular}
\caption{Performance Comparison. BLOCKIE-Sonnet outperforms the state-of-the-art across all three datasets}
\label{tab:performance}
\end{small}
\vskip -0.1in
\end{table*}

\subsubsection{BLOCKIE helps smaller LLMs outperform large LLMs}
 We examine BLOCKIE's robustness to LLMs by evaluating performance across LLMs of varying sizes. As shown in Table \ref{tab:llmstability}, BLOCKIE maintains strong performance even with smaller models - BLOCKIE with Qwen 2.5 32B (96.14\% F1) outperforms LMDX-Gemini Pro (~200B parameters, 95.57\% F1) and Sonnet Zero-Shot as well as Few-shot (91.37\% and 95.72\% respectively), while BLOCKIE with Qwen 2.5 7B (87.72\% F1) significantly surpasses other approaches using similar-sized models like DocLLM (67.4\% F1) and LayoutLLM (63.1\% F1). Note that the finetuned version of the Qwen 32B model falls short of Sonnet Few shot significantly (91.08\% vs 95.72\%), showing that the improvement in performance is caused by BLOCKIE and not purely the abilities of the LLM. 

\begin{table}[ht]
\vskip 0.15in
\begin{center}
\begin{small}
\begin{sc}
\begin{tabular}{ll}
\toprule
                & CORD - SER \\
Approach                &            \\
\midrule
DocLLM - 7B             & 67.4       \\
LayoutLLM - 7B          & 63.1       \\
LMDX - Gemini Pro       & 95.57      \\
\midrule
Qwen 2.5 7B Finetuned  & 84.03      \\
Qwen 2.5 14B Finetuned & 89.36      \\
Qwen 2.5 32B Finetuned & 91.08      \\
Sonnet - Zero shot & 91.37      \\
Sonnet - Few shot & 95.72      \\
BLOCKIE - Qwen 2.5 7B  & 87.72      \\
BLOCKIE - Qwen 2.5 14B & 89.98      \\
BLOCKIE - Qwen 2.5 32B & \textbf{96.14}      \\
BLOCKIE - Qwen 2.5 72B & \textbf{96.01}      \\
BLOCKIE - Sonnet 3.5   & \textbf{98.83}     
\end{tabular}
\caption{BLOCKIE with smaller LLMs outperforms massive state-of-the-art models Sonnet and Gemini Pro}

\label{tab:llmstability}
\end{sc}

\vskip -0.1in
\end{small}

\end{center}
\end{table}

\subsubsection{BLOCKIE is resistant to heterogeneity and to unseen document formats.}
To assess format resilience, we conduct two experiments. In the first experiment, we evaluate performance when training on only 100 samples selected for maximum format diversity (based on maximising text embedding distances with the test sample). Table \ref{tab:hundred} shows that while LayoutLMV3's performance drops significantly from 96.98\% to 78.79\% with diverse samples, BLOCKIE maintains robust performance (94.47\% F1), demonstrating better generalization to format variations. This is even better that 91.48\% achieved by \citealp{perot2024lmdxlanguagemodelbaseddocument} by training on 100 random samples.

In our second experiment, we evaluate cross-dataset generalization by testing a CORD-trained model on SROIE documents (using the enity total amount, which is common in both datasets). As shown in Table \ref{tab:hundred}, BLOCKIE maintains strong performance (97.06\% F1) while LayoutLMV3's performance deteriorates substantially (33.43\% F1), further validating our approach's resilience to format changes.
\begin{table*}[t]
\vskip 0.15in
\begin{center}
\begin{small}
\begin{sc}
\begin{tabular}{lll}
\toprule
 Test on               & CORD - SER  & SROIE -Total Amount \\
                \midrule
Trained on                &  [100 train samples  &  [train samples from CORD] \\

                &  least similar to test]  &   \\
\midrule
LayoutLMV3 & 78.79 & 33.43   \\
Sonnet 3.5 Few Shot & 92.11 &  95.39     \\
BLOCKIE - Qwen 2.5 32B & 86.51  & 91.01     \\
BLOCKIE - Sonnet 3.5   &\textbf{ 94.47}    & \textbf{97.06} \\  
\end{tabular}
\caption{Resilience to heterogeneity and new formats. Sonnet is more resilient than LayoutLMV3, and BLOCKIE further enhances this resilience, outperforming layout-aware NLP methods designed to recognize templates.}
\label{tab:hundred}
\end{sc}

\vskip -0.1in

\end{small}
\end{center}
\end{table*}
\subsubsection{Block creation is crucial for BLOCKIE performance}
\label{blockperf}
The effectiveness of BLOCKIE relies critically on accurate semantic block creation. Our analysis reveals that block creation quality strongly correlates with final extraction performance (Table \ref{tab:blockcreate}). The performance gap between different model sizes can be largely attributed to their block creation capabilities - Qwen 32B and 72B achieve state-of-the-art performance due to superior block creation (85.03\% and 81.69\% block-level F1\footnote{Block level F1 is derived by comparison with ground truth blocks created using labelled data} respectively), while smaller models show lower block creation accuracy.
\begin{table}[t]
\vskip 0.15in
\begin{center}
\begin{small}
\begin{sc}
\begin{tabular}{lll}
\toprule
                & CORD - SER & \\
Approach                &      Block F1  &      Entity  F1      \\
\midrule
BLOCKIE - Qwen 2.5 7B  & 74.91 & 87.72      \\
BLOCKIE - Qwen 2.5 14B & 73.25 & 89.98      \\
BLOCKIE - Qwen 2.5 32B & 85.03& 96.14      \\
BLOCKIE - Qwen 2.5 72B & 81.69 & 96.01      \\
BLOCKIE - Sonnet 3.5   & 86.73 & 98.83     
\end{tabular}

\caption{Correlation between block creation accuracy and performance.}
\label{tab:blockcreate}
\end{sc}

\vskip -0.1in
\end{small}

\end{center}
\end{table}

To isolate the impact of block creation, we evaluate smaller models (7B, 14B) using ground truth blocks and blocks created by the 32B model. As shown in Table \ref{tab:blockperf}, with perfect blocks, even 7B and 14B models achieve performance comparable to larger models (94.38\% and 94.98\% F1 respectively), indicating that block creation quality is the primary performance bottleneck.

\begin{table}[t]
\vskip 0.15in
\begin{center}
\begin{small}
\begin{sc}
\begin{tabular}{llll}
\toprule
BLOCKIE                 &  End   &  Qwen 32B   &     Ground Truth  \\
Qwen Size          &   To End &   blocks  &    Blocks  \\
\midrule
7B  & 87.72  & 90.91 & 94.38     \\
14B & 89.98& 92.23   & 94.98      \\

\end{tabular}
\caption{Semantic Block F1-scores. After correcting semantic blocks of test samples, smaller models are able to recover 70\% of the 10 percent performance gap with larger models}
\label{tab:blockperf}
\end{sc}

\vskip -0.1in
\end{small}

\end{center}
\end{table}

\subsubsection{BLOCKIE is able to perform value-absent inference}
Finally, we demonstrate BLOCKIE's reasoning capabilities through value-absent inference. We evaluate on CORD receipts where line item counts are not explicitly stated but can be inferred through counting. On a sample of 20 such cases, BLOCKIE successfully infers the correct count in 18 instances (90\% accuracy), handling complex scenarios including implicit quantities and hierarchical items. Figure \ref{quantity_infer} illustrates several challenging cases where BLOCKIE successfully performs multi-step reasoning to arrive at correct inferences. This capability distinguishes BLOCKIE from existing approaches that are limited to extracting explicitly present information.

\label{quantity_infer}
\begin{figure}[ht]
\vskip 0.2in
\begin{center}
\centerline{\includegraphics[width=\columnwidth]{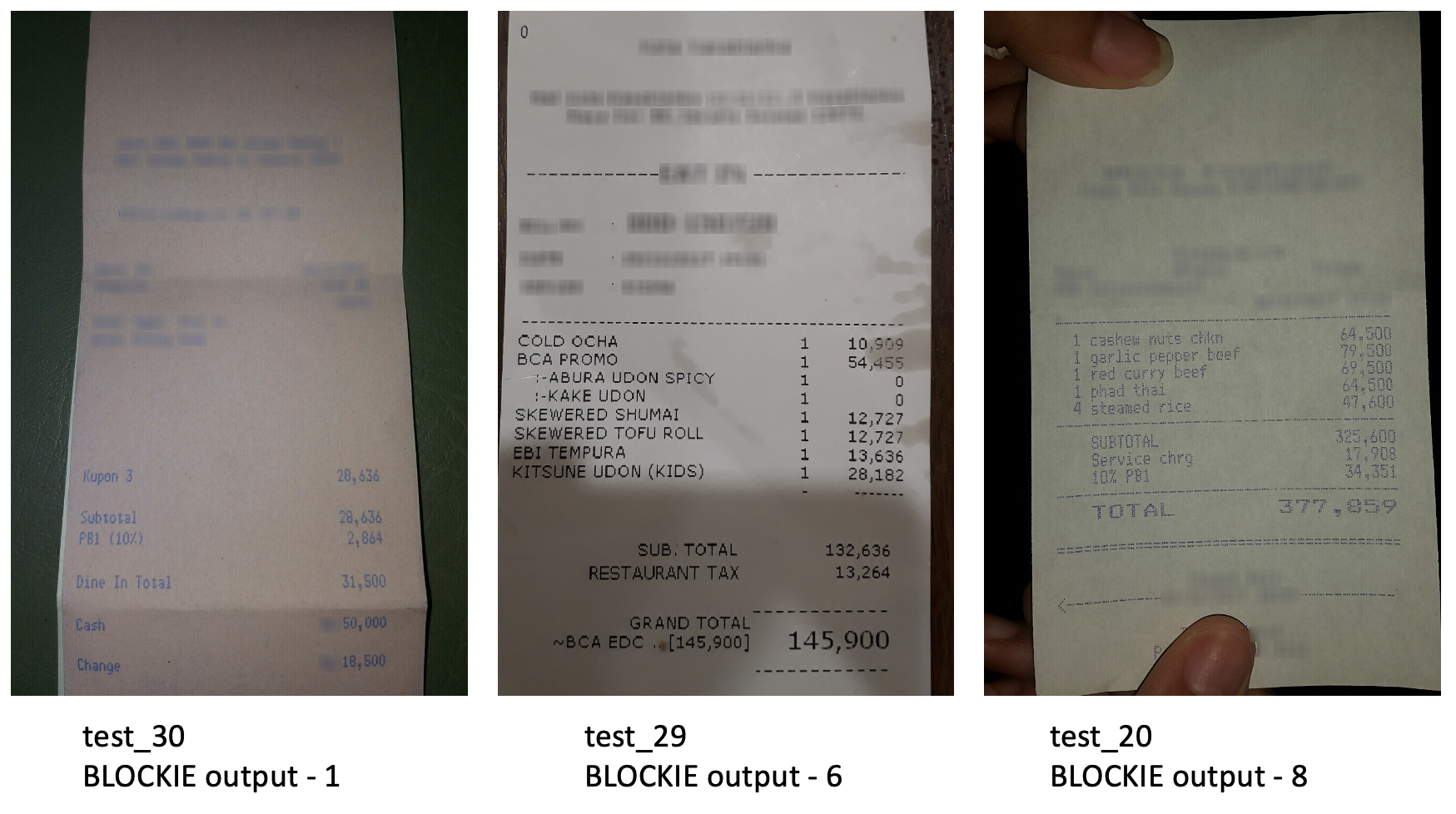}}
\caption{Some challenging inferences made by BLOCKIE. In test\_30, the single line item does not have a quantity mentioned. In test\_29, the LLM has to reason to leave out sub-items from the count. In test\_20, it has to perform a multi-step addition.}
\label{blocks}
\end{center}
\vskip -0.2in
\end{figure}

\section{Conclusion}
In this work, we introduced the concept of \textbf{semantic blocks} and proposed a novel LLM-based approach for information extraction from documents leveraging them. The segmentation of documents into generalizable, smaller, self-contained semantic blocks allowed LLMs to generate focused step-by-step reasoning guiding their annotation, and we demonstrated that this was effective by showing state-of-the-art performance across diverse public datasets. 

The framework is designed to be generalizable across various large language models (LLMs) and resilient to unseen document layouts and formats, and we demonstrated robust performance across multiple LLMs, heterogeneity and new, unseen document formats. Additionally, we also showcased the ability of BLOCKIE to perform value-absent inference.

The combination of semantic reasoning, robust generalization, and resilience to variation positions this methodology as a promising direction for future research in document information extraction. Future work could focus on incorporating image-based features such as font size, qualities such as bold/italics, etc, into semantic block creation even in text-only LLMs.

\section*{Limitations}
We acknowledge the limitations of BLOCKIE with a view to motivating further research in this field. The computational architecture currently requires sequential LLM calls for block creation, processing and combining which increases latency. While our block creation methodology showed robust performance across all three datasets and experiments, it could be refined further. Specifically, the current block creation methodology does not leverage image-based contextual clues such as font, italics/bold, visual markers for linkages such as arrows, etc. Additionally, while robust performance was observed across 5 different LLMs of varying sizes, BLOCKIE's performance is inherently tied to the reasoning capability of the LLM being used. As was shown in  section \ref{blockperf}, it is vital to ensure that the LLM is able to reason and create proper blocks with linked semantic atoms, as missed linkages can be hard to recover. Future research should focus on robust block creation using the definition of semantic blocks and linked semantic atoms. Finally, using proprietary LLMs like Sonnet can make BLOCKIE less transparent even with step-by-step reasoning output, and caution needs to be exercised to ensure outputs are as expected.
\bibliography{custom2}

\appendix

\section{BLOCKIE prompt templates}
\label{sec:appendix1}

\begin{table}[ht]
\caption{Train Dataset Labeling Prompt Template}
\vskip 0.15in
\begin{center}
\begin{small}
\begin{sc}
\begin{tabular}{|p{0.95\columnwidth}|}
\hline
\textbf{Prompt Instructions} \\
\hline
Take the following text - \texttt{<text>} \\
This gets parsed into \texttt{<annotation>} \\
Break the provided text into semantic blocks \\
like $block_1,block_2$... with \textbf{related text} \\
in same block. Here are some rules: \\
1/ The output should be a dictionary with \\
keys - $block\_1, block\_2$ etc. \\
2/ Each block should be a dictionary itself, \\
with the keys - reason, text and parsed: \\
\hspace{1em}• In Reason, think step-by-step why \\
\hspace{2em}the text under consideration is a \\
\hspace{2em}single block \\
\hspace{1em}• The text key should contain the \\
\hspace{2em}text present in the block \\
\hspace{1em}• The parsed section should contain \\
\hspace{2em}the part of the parsed output \\
\hspace{2em}the text maps to \\
3/ Related text refers to text belonging to \\
the same  <linked or hierarchical entity from schema, or others> \\
4/ Do not leave out any text \\
5/ Do not write a single extra word \\
\hline
\end{tabular}
\end{sc}
\end{small}
\end{center}
\vskip -0.1in
\label{tab:labeling}
\end{table}

\clearpage

\begin{table}[ht]
\caption{Block Creator Prompt Template}
\vskip 0.15in
\begin{center}
\begin{small}
\begin{sc}
\begin{tabular}{|p{0.95\columnwidth}|}
\hline
\textbf{Parser Instructions} \\
\hline
You are a seasoned text parser. Given an \\
OCR text, you are able to parse it into \\
blocks of related text along with \\
step-by-step reasons. \\
\\
\texttt{<Linked and Hierarchical entity} \\
\texttt{identification rules>}: \\
\texttt{<Few Shot Examples>} - \\
Here are some rules: \\
1/ The output should be a dictionary with \\
keys - block\_1, block\_2 etc. \\
2/ Each block should be a dictionary itself, \\
with the keys - reason, and text. \\
\hspace{1em}a. In Reason, think step-by-step why \\
\hspace{2em}the text under consideration is a \\
\hspace{2em}single block. Show step by step \\
\hspace{2em}reasoning using rules and examples \\
\hspace{2em}laid out. \\
\hspace{1em}b. The text key should contain the \\
\hspace{2em}text present in the block. \\
3/ Related text refers to text belonging \\
to the same <linked or hierarchical entity from schema, or others> \\
4/ Do not leave out any text. \\
5/ Do not write a single extra word. \\
\\
\texttt{<Verification Process>} \\
Complete the Answer for the following text. \\
Do not write anything extra. \\
\texttt{<OCR words>} \texttt{<bounding boxes>} \\
Answer: \\
\hline
\end{tabular}
\end{sc}
\end{small}
\end{center}
\vskip -0.1in
\label{tab:parser}
\end{table}

\begin{table}[ht]
\caption{Block Parser Prompt Template}
\vskip 0.15in
\begin{center}
\begin{small}
\begin{sc}
\begin{tabular}{|p{0.95\columnwidth}|}
\hline
\textbf{System Instructions} \\
\hline
You are an expert system for parsing receipt \\
text blocks into structured data. Your role \\
is to analyze receipt text and convert it \\
into a structured dictionary format. \\
\\
\texttt{<SCHEMA AND FIELD DESCRIPTIONS>} \\
\texttt{<Formatting rules>} \\
\\
SIMILAR EXAMPLES FOR REFERENCE: \\
\texttt{<few shot examples>} \\
Note: These examples are for reference but \\
may contain some inconsistencies. Follow \\
the rules above strictly. \\
\\
\textbf{CURRENT TASK:} \\
This is a block created previously where \\
the block-creator had this reason \\
"\texttt{\{query\_reason\}}" \\
\\
Your task is to create a complete, valid \\
JSON dictionary following the provided \\
schema that represents all the information \\
in this receipt document. \\
\\
\texttt{<OUTPUT SPECIFICATION>} \\
\texttt{<Verification Process>} \\
\\
Parse this receipt block into the schema \\
format: \\
\texttt{<query\_block>} \\
\hline
\end{tabular}
\end{sc}
\end{small}
\end{center}
\vskip -0.1in
\label{tab:receipt-parser}
\end{table}

\clearpage

\begin{table*}[ht]
\caption{Block Combiner Prompt Template}
\vskip 0.15in
\begin{center}
\begin{small}
\begin{sc}
\begin{tabular}{|p{0.95\columnwidth}|}
\hline
\textbf{System Instructions} \\
\hline
You are an expert system for parsing receipt \\
documents into structured data. Your task is \\
to analyze a complete receipt document and \\
create a comprehensive dictionary using \\
partial information from individual blocks. \\
\\
\textbf{CONTEXT:} \\
You will be provided with: \\
1. All the words in the document \\
2. Bounding boxes \\
3. Individual blocks of text and their \\
\hspace{1em}partial parses \\
4. The required dictionary schema \\
\\
\texttt{<SCHEMA AND FIELD DESCRIPTIONS>} \\
\texttt{<Linked and Hierarchical entity} \\
\texttt{identification rules>} \\
\\
\textbf{ALL WORDS IN THE DOCUMENT:} \\
\texttt{\{text\}} \\
\\
\textbf{ALL BOUNDING BOXES IN THE DOCUMENT:} \\
\texttt{\{bboxes\}} \\
\\
\textbf{PARSED BLOCKS:} \\
Below are the individual blocks and their \\
partial parses along with reason. Use these \\
to help construct the complete dictionary: \\
\texttt{\{blocks\_and\_parses\}} \\
\\
\textbf{INSTRUCTIONS:} \\
1. Use the complete document text to \\
\hspace{1em}understand the full context \\
2. Utilize the partial parses from blocks \\
\hspace{1em}to help construct the final dictionary \\
\hspace{1em}- remember - the partial parses may \\
\hspace{1em}not have full context \\
3. Ensure all information is correctly \\
\hspace{1em}categorized according to the schema \\
4. Maintain consistency with numerical \\
\hspace{1em}formats from the original text \\
\\
\texttt{<Verification Process>} \\
\\
Your final dictionary should contain two \\
keys: \\
1. reason - justify step by step why you \\
\hspace{1em}chose particular values. Use the \\
\hspace{1em}reason from partial parses, check if \\
\hspace{1em}it mentions exact match. \\
2. invoice - share the invoice dictionary \\
\\
Return only the final JSON dictionary \\
without any additional explanation with \\
proper format. \\
\hline
\end{tabular}
\end{sc}
\end{small}
\end{center}
\vskip -0.1in
\label{tab:doc-parser}
\end{table*}

\clearpage

\section{Datasets and Benchmarks}
\label{sec:appendix2}

\subsection{Datasets}
\textbf{CORD Dataset}
CORD \cite{park2019cord} contains 1000 Indonesian receipts, divided into train, validation and test samples of size 800,100 and 100. Along with the images, CORD also contains crowdsourced labels, and OCR output with bounding boxes. 30 hierarchical entities are annotated manually under top-level entities menu, subtotal and total. The associated task is to assign the words in the OCR output to these entities. Performance is assessed using micro-F1 on entity prediction.

\textbf{SROIE Dataset}
SROIE \cite{Huang_2019} dataset consists of scanned receipts from a variety of domains, such as retail, food, and services, split into 626 train and 347 test receipts. The dataset contains images, OCR output and annotations with labeled entities for Company Name, Date, Total Amount, and Address. We evaluate our approach on the information extraction task proposed in the paper. Performance is assessed using micro-F1 on entity prediction.

\textbf{FUNSD Dataset}
The FUNSD Datatet \cite{funsd} contains 199 fully annotated images of forms sampled from the form type document of the RVL-CDIP dataset \citep{harley2015icdar}. The dataset is split into 149 images in the training set and 50 in the testing set. The annotations consist of text with four keys - question, answer, header, and others, which is simplistic and do not represent meaningful entities. However, the annotations also contain linkages, forming meaningful question-answer pairs and groupings of these pairs under headers. We focus on the entity-linking task to evaluate the ability of our approach to extract meaningful relations. 

\subsection{LLMs and Benchmark approaches}
We tested out BLOCKIE across 5 different LLMs from two different families. The LLMs chosen are widely used and vary in sizes from massive proprietary models to open-source models with 7B parameters.

\textbf{Claude 3.5 Sonnet} 
Claude 3.5 Sonnet is the first model released by Anthropic from the Claude 3.5 family \cite{Anthropic35}. In the benchmark evaluations released by Anthropic, it showed at-par or superior performance compared to Claude 3 Opus, the previous best-performing Anthropic model, while being 2x faster. It established new state-of-the-art on reasoning and question-answering tasks at the time of its release.

For few-shot Sonnet results, we conducted experiments using the CORD validation dataset and found best results when 5 examples were used that were the closest (with respect to text embedding similarity) to the target sample.

\textbf{Qwen 2.5} 
Qwen 2.5 is a family of open-source LLMs released by Alibaba Cloud \cite{qwen2025qwen25technicalreport}. The family contains both base language models, instruction-tuned models as well as specialized models for coding, math, etc. The family consists of models in sizes varying from 0.5B parameters to 32B parameters. We used the 72B, 32B, 14B and 7B versions for our experimentation. 

For finetuning, we used LORA \citep{lora} with rank 64 for 6 epochs with learning rate 0.00002. These numbers were based on results obtained on the validation dataset of CORD. 

\textbf{LayoutLMV3} 
LayoutLMV3 \cite{layoutlmv3} is a state-of-the-art information extraction benchmark. It incorporates layout information using cross-attention between bounding boxes and text, and through masked image modeling. It shows competitive performance on all three benchmark datasets. Note that while \cite{cvpr2023geolayoutlm} outperforms LayoutLMV3, the authors have not officially released their pre-processing code or fine-tuned weights for CORD. We use layoutlmv3 in our experiments to demonstrate the limitations of SER-based approaches.

When we finetuned LayoutLMV3 for our experiments on heterogeneity and value-absent inference, we used the parameters listed in the official paper for CORD.

We reviewed the licenses for all these datasets and models, and ensured that we stick to the intended usage of these for research purposes.
\end{document}